\documentclass[prl,twocolumn, showpacs,preprintnumbers,amsmath,amssymb,superscriptaddress]{revtex4}

\usepackage{graphicx}
\usepackage{dcolumn}
\usepackage{bm}

\begin{document}

\preprint{APS/123-QED}

\title{Muon Spin Relaxation and Susceptibility Studies of Pure and 
Doped Spin 1/2 Kagom\'{e}-like 
system (Cu$_x$Zn$_{1-x}$)$_{3}$V$_{2}$O$_7$(OH)$_{2}$ 2H$_2$O}

\author{A.~Fukaya}
 \altaffiliation[Present address]{
Institute of Materials Research,
Tohoku University,
Katahira, Aoba-ku, Sendai, 980-8577, 
Japan}
\affiliation{Physics Department, Columbia University, 538 West, 
120th Street, New York, NY 10027, USA\\}%
\author{Y.~Fudamoto}
\affiliation{Physics Department, Columbia University, 
538 West, 120th Street, New York, NY 10027, USA\\}%
\author{I.M.~Gat}
\affiliation{Physics Department, Columbia University, 538 West, 
120th Street, New York, NY 10027, USA\\}%
\author{T.~Ito}
\affiliation{Physics Department, Columbia University, 538 West, 
120th Street, New York, NY 10027, USA\\}%
\affiliation{Correlated Electron Research Center (CERC), AIST, 
Tsukuba, Ibaraki 305-8562, Japan\\}%
\author{M.I.~Larkin}
\affiliation{Physics Department, Columbia University, 
538 West, 120th Street, New York, NY 10027, USA\\}%
\author{A.T.~Savici}
\affiliation{Physics Department, Columbia University, 
538 West, 120th Street, New York, NY 10027, USA\\}%
\author{Y.J.~Uemura}
\affiliation{Physics Department, Columbia University, 
538 West, 120th Street, New York, NY 10027, USA\\}%
\author{P.P.~Kyriakou}
\affiliation{Department of Physics and Astronomy, 
McMaster University, Hamilton, Ontario L8S 4M1, Canada\\}%
\author{G.M.~Luke}
\affiliation{Department of Physics and Astronomy, 
McMaster University, Hamilton, Ontario L8S 4M1, Canada\\}%
\author{M.T.~Rovers}
\affiliation{Department of Physics and Astronomy, 
McMaster University, Hamilton, Ontario L8S 4M1, Canada\\}%
\author{K.M.~Kojima}
\affiliation{Department of Frontier 
Sciences, The University of Tokyo, Hongo, Bunkyo, Tokyo 114-8656, Japan\\}%
\author{A.~Keren}
\affiliation{Department of Physics, 
Technion-Israel Institute of Technnology, Haifa 32000, Israel\\}%
\author{M.~Hanawa}
\affiliation{Institute for Solid State Physics, 
University of Tokyo, Kashiwanoha, Kashiwa, Chiba 277-8581, Japan\\}%
\author{Z.~Hiroi}
\affiliation{Institute for Solid State Physics, 
University of Tokyo, Kashiwanoha, Kashiwa, Chiba 277-8581, Japan\\}%

\date{\today}

\begin{abstract}  
Muon spin relaxation ($\mu$SR) and magnetic susceptibility measurements 
have been performed on the pure and diluted spin 1/2 kagom\'{e} 
system (Cu$_x$Zn$_{1-x}$)$_{3}$V$_{2}$O$_7$(OH)$_{2}$ 2H$_2$O. 
In the pure $x=1$ system we found 
a slowing down of Cu spin fluctuations with decreasing temperature towards
$T \sim 1$ K, followed by slow and nearly temperature-independent  
spin fluctuations persisting down to $T$ = 50 mK, indicative of
quantum fluctuations.
No indication of static spin freezing was 
detected in either of the pure ($x$=1.0) or diluted samples.     
The observed magnitude of 
fluctuating fields indicates that  
the slow spin fluctuations 
represent an intrinsic property of kagom\'e network rather
than impurity spins.  
\end{abstract}
%

\pacs{Valid PACS appear here}
\maketitle

\newcommand{\CVO}{Cu$_{3}$V$_{2}$O$_7$(OH)$_{2}$ 2H$_2$O}
\newcommand{\CZV}{(Cu$_x$Zn$_{1-x}$)$_{3}$V$_{2}$O$_7$(OH)$_{2}$ 2H$_2$O}
\newcommand{\mSR}{$\mu$SR}
\newcommand{\LF}{$H_{\rm LF}$}
\newcommand{\Hloc}{$H_{\rm loc}$}
\newcommand{\Hlocd}{$H'_{\rm loc}$}

\parskip 0cm

Geometrically frustrated interactions bring new types of 
cooperative phenomena to spin systems.  Spins on the kagom\'{e} lattice, 
coupled with  antiferromagnetic nearest neighbor interactions, 
are known as a prototype of strongly frustrated 
systems~\cite{Ramirez94,Schiffer96}.  
Theoretical and experimental studies of kagom\'{e} systems 
have been extensively performed.  Materials studied as 
candidates of kagom\'{e} system~\cite{Ramirez94,Schiffer96}
include SrCr$_{9p}$Ga$_{12-9p}$O$_{19}$ (SCGO) and
jarosite family AM$_3$(OH)$_6$(SO$_4$)$_2$ 
(A: typically a univalent, M=Fe$^{3+}$,Cr$^{3+}$).  
Some of them (e.g. Fe-jarosite) undergo a transition to 
long-range N\'{e}el order, while others (SCGO and Cr-jarosite:
both with $S$=3/2 Cr moments)
exhibit spin-glass-like behavior in susceptibility. 
 
Quantum effects should play important roles in the 
magnetism of kagom\'{e} compounds, similar to the 
case of low dimensional spin systems.  
A theoretical study of 
$S$=1/2 Heisenberg kagom\'{e} systems 
indicates that the ground state 
is spin singlet~\cite{theory2}.  Ramirez {\it et al.\/}~\cite{SCGO_C} 
reported independence of the specific heat of $S$=3/2 kagom\'{e} system SCGO 
on high applied magtic fields, and interpreted this in terms of  
low-energy excitations dominated by spin singlet states~\cite{SCGO_C,Sind00}.
On the other hand, muon spin relaxation ($\mu$SR) results in 
SCGO~\cite{Uemura94} suggest that dynamic spin fluctuations persist even at 
$T \rightarrow 0$.  Experimental studies of $S$=1/2 kagom\'{e}
systems would be useful for better understanding of quantum effects.     
No good candidate of $S$=1/2 kagom\'{e} material, however, has been 
found until recently.

Hiroi {\it et al.\/} found that $S$=1/2 Cu$^{2+}$ spins in 
the copper
volborthite \CVO \ (CVO) forms a spin network, which is only
slightly distorted from a complete kagom\'{e} lattice~\cite{Hiroi01}.
Vanadium is in a non-magnetic state, as confirmed by NMR~\cite{Hiroi01}.  
Neither transition to long-range order nor a spin-glass like transition 
was observed down to 1.8~K in magnetization,
heat-capacity and NMR measurements.  An ESR study suggests existence of 
short-range spin correlations below 5~K~\cite{Okubo01}. 
In order to characterize ground state of CVO, we have performed 
muon spin relaxation (\mSR) measurements down to 50~mK.

Effects of magnetic dilution in kagom\'{e} systems are also interesting.
We could possibly obtain insights both about doped and undoped systems,
as demonstrated in the case of SCGO~\cite{Martinez92,Keren00}.   
Thus, we attempted to substitute Cu$^{2+}$ ions by non-magnetic 
Zn$^{2+}$ ions. Zn$_{3}$V$_{2}$O$_7$(OH)$_{2}$ 2H$_2$O, with the same 
stoichiometry~\cite{Zavalij}, has a crystal structure  
slightly different from that of CVO. Both the 
lattice parameter and atom positions are, 
however, quite similar between the Cu and Zn systems. 
We have succeeded in the synthesis of mixed compound 
\CZV. In this letter, we report susceptibility 
($\chi$) and \mSR\ measurements 
of pure and diluted CVO and discuss some exotic features 
of their ground states.


Measurements of $\mu$SR and $\chi$ were performed in five 
powder samples with nominal $x$=1.0, 0.95, 0.90, 0.70 and 0.60.
Powder samples of pure CVO ($x$ = 1.0) 
were prepared in the method described in 
Ref.~\onlinecite{Hiroi01}. The diluted samples were also made in a  
similar method, but using  
CuSO$_4$$\cdot$5H$_2$O, ZnSO$_4$ 5H$_2$O, V$_2$O$_5$ and NaOH as 
starting materials. 
X-ray powder diffraction patterns of these samples indicated they were single phase. 
DC susceptibility $\chi$ at $T > 1.8$ K was measured with 
a SQUID magnetometer at Columbia, while
$\mu$SR measurements at 50 mK $<$ T $<$ 6 K were performed at TRIUMF. 
To assure cooling of 
samples used for $\mu$SR 
in a dilution refrigerator, we mixed the samples 
with 20~\% wt. of Au powder and pressed them into pellets.



\begin{figure}[t]
\input{epsf}
\epsfxsize 6.5cm
\centerline{\epsfbox{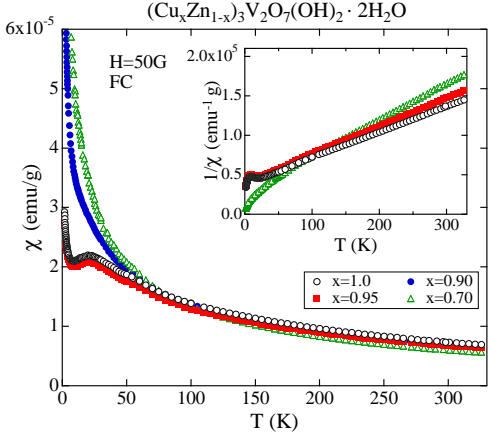}}
\caption{(a) 
Magnetic susceptibility of CVO with various Cu concentrations 
in $H$=50~G in the field-cooled condition. The inset shows the inverse 
susceptibility of $x$=1.0, 0.95 and 0.90.} 
\label{fig:XT}
\end{figure}
In Fig.~1(a), we plot the susceptibility $\chi(T)$, 
as a function of temperature $T$, measured in a magnetic field of 50~G.  
Only the field-cooled results are shown, since we did not
observe any hysteresis. 
In $x$=1.0, $\chi(T)$
shows a broad maximum at $\sim$20~K. 
Below 9 K, the susceptibility begins to increase
with decreasing $T$. This increase indicates a Curie-like term due to 
impurities or lattice defects. In $x$=0.95, the behavior is quite similar 
to that in $x$=1.0.  With increasing dilution, the increase of $\chi$ 
at low temperature becomes more pronounced, as expected for 
un-paired Cu moments created by Zn.  The broad maximum at 20 K was
no longer observed for $x \leq 0.9$.  In any of these samples, we 
did not find any signature 
indicating either a transition to long-range order or spin freezing 
at $T > 1.8$K.  We found no anomaly in $\chi$ 
at the percolation threshold, $p_{\rm c}=0.65$~\cite{Essam72}.

The inverse susceptibility for $x$=1.0, 0.95 and 0.70 is shown in 
the inset of Fig.~1. In all concentrations, $1/\chi$ shows nearly  
linear variation with $T$ down to $\sim$100~K, though 
it slightly deviates from linearity below $\sim$200~K.
The Weiss temperature $\Theta$ estimated from 1/$\chi$ above 200 K   
in the pure $x=1$ compound was $-117$~K, consistent with 
Hiroi {\it et al.\/}~\cite{Hiroi01}, who estimated 
the exchange interaction $J$ = 84 K from $\chi$.
The absolute value of $\Theta$ decreases with decreasing Cu concentration.

%
\begin{figure}
\label{fig:spectra}
\input{epsf}
\epsfxsize 8.0cm
\centerline{\epsfbox{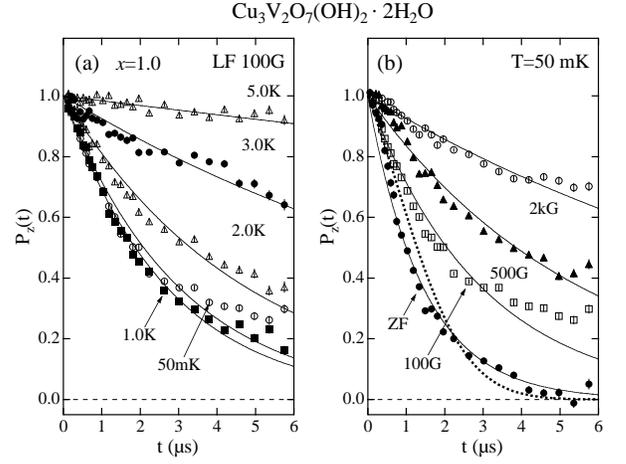}}
\caption{ (a) Temperature dependence of the time spectra of 
the $x$=1.0 sample measured in \LF=100~G. (b) Longitudinal 
field dependence of the time spectra of the $x$=1.0 sample 
at $T$=50~mK. The solid lines show the 
results of fitting with Eq. (1). The broken line in (b) shows the spectrum
at $T$= 50 mK in \LF = 100 G multiplied to ZF spectrum at $T$ = 6 K
(which represents the effect of NDF).} 
\end{figure}

\begin{figure}
\label{fig:mSRe}
\input{epsf}
\epsfxsize 7.5cm
\centerline{\epsfbox{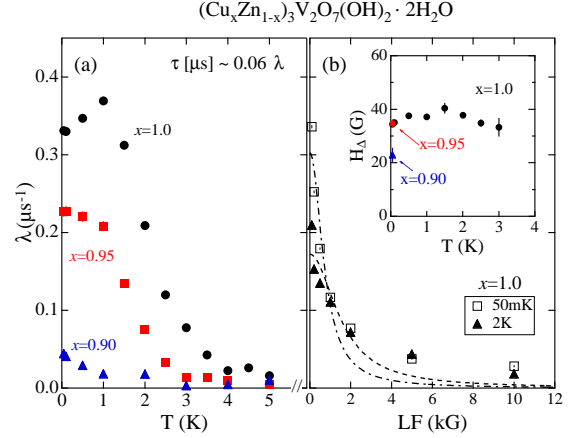}}
\caption{(a) Temperature dependence of the 
relaxation rate $\lambda$ 
of the $x$=1.0, 0.95, and 0.90 sample
in \LF = 100 G. For the instantaneous local field
width $H_{\Delta}\sim 35$G, the correlation time becomes 
$\tau$ [$\mu$s] $\sim 0.06 \lambda$ [/$\mu$s]. 
 (b) Longitudinal field 
dependence of $\lambda$ at $T$=50~mK and 2.0~K. 
The dotted curves show the results of fitting with Eq. (2).
The inset in (b) shows the temperature 
dependence of the field amplitude $H_{\Delta}$.} 
\end{figure}
To characterize static/dynamic spin correlations 
with a microscopic probe, 
we performed \mSR \ measurements. Figure 2(a) 
shows the 
time spectra of muon polarization, $P_{\rm z}(t)$ obtained,
after subtraction of a temperature-independent background/Au contribution, 
for $x$=1.0 in a longitudinal field (LF) of \LF = 100~G 
applied to decouple nuclear dipolar fields (NDF).
The Kubo-Toyabe NDF width~\cite{YJU85} $\Delta$ = 0.39 /$\mu$s 
was estimated from the spectrum at $T$ = 6 K in zero-field (ZF). 
As $T$ decreases, the relaxation rate in Fig. 2(a) increases down 
to $\sim$1~K, followed by a saturation at lower 
temperatures.  

The time spectra at 50~mK in ZF and some selected LF
are shown in Fig.~2(b). 
We did not 
observe oscillation of muon spin in ZF.
The persisting relaxation in \LF $>$ 100 G clearly indicates
that the observed relaxation at $T$ = 50 mK is predominantly due to 
fluctuating local fields from Cu moments.  In such a case, one
woule expect the ZF spectra to be given by a product of 
a low-field LF spectrum and a Kubo-Toyabe decay due to NDF~\cite{YJU85},
which is shown by the dotted line in Fig. 2(b).  
A reasoable agreement of this line and the ZF data 
further confirms the dynamic origin of the relaxation in LF. 

In a further inspection of the line shape, we notice that
(1) the early-time decay in \LF=100 and 500 
G is somewhat rounder than the exponential shape (solid lines); 
(2) the ZF spectrum shows slight deviation from the broken line; and
(3) the longer time decay after $t$ = 4 $\mu$s in LF becomes
slower than exponential.
These features have been found in 
the so-called ``undecouplable Gaussian''
$\mu$SR line shapes, observed in kagom\'{e} systems, 
SCGO~\cite{Uemura94} and Cr-jarosite~\cite{Keren96}, 
charge-doped Haldane gap system (Y,Ca)$_2$BaNiO$_5$~\cite{Kojima95}, 
body-centered tetragonal spin system CePt$_{2}$Sn$_{2}$~\cite{Luke97}, 
and a dimer spin-gap system SrCu$_2$(BO$_{3}$)$_{2}$~\cite{Fukaya03}. 
The origin of this anomalous behavior has not yet been clearly understood.
In the present CVO system, however, this feature 
appears in a much less prominent way than in the above-mentioned systems.
Therefore, to grasp dominant trends of the data, 
we fitted the LF time spectra at $t < 4 \mu$s 
with a simple exponential function, 
\begin{equation}\label{eq:rlxfunc} 
{\rm P_z}(t)=\exp(-\lambda t),
\end{equation}
where $\lambda$ is the relaxation rate. 
The results of this fit are shown 
by the solid lines in Fig.~2.

The temperature dependence of $\lambda$ is shown 
in Fig.~3(a).  With decreasing
temperature, $\lambda$ starts to increase 
at $T\sim 4$ K and shows a saturation around 
$T$ = 1.0~K in $x$ = 1.0 and 0.95. 
The LF dependence of $\lambda$ in $x$ = 1.0 is
shown in Fig.~3(b). 
For a Markovian correlation of 
local fields characterized by 
a Gaussian distribution $P(H_{i}) \propto
\exp(-H_{i}^{2}/2H_{\Delta}^{2})$ ($i$ = x,y,z) with the amplitude of 
$H_{\Delta}$ and 
the correlation time $\tau$, 
$\lambda$ follows a Lorentzian function of \LF,
\begin{equation}
\lambda (H_{\rm LF}) =  
\frac{2\gamma_{\rm \mu}^2 H_{\Delta}^2 \tau}
{1 + \gamma_{\rm \mu}^2 H_{LF}^2 \tau^2 },
\end{equation}
where $\gamma_{\rm \mu}$ ($\gamma_{\rm \mu} =2\pi \times 
13.55 \times 10^3$ Hz/G) 
is the gyromagnetic ratio of muon~\cite{YJU85}. 
The dotted lines in Fig.~3(b) show a fit to this function. 
The observed LF dependence significantly deviates from
Lorentzian.  
This may suggest a wide distribution of $H_{\Delta}$ and/or $\tau$, 
or non-exponential / non-Markovian type of time correlations of Cu spins. 

For a Lorentzian power function of Eq.~(2), $1/\tau$ 
can be derived from the half width at 
half maximum in the $\lambda$ vs LF curve. Thus, assuming that the 
value of $\lambda$ in 100~G is equal to that in zero-field, we 
obtained the value of \LF\ where $\lambda$ becomes half of that in 100~G,
multiplied it with $\gamma_{\rm \mu}$, and estimated  
the characteristic effective fluctuation rate $1/\tau$. We also derive the  
corresponding amplitude $H_{\Delta}$ of 
the instantaneous local fields from the 
values of $\lambda$ in 100~G 
using Eq.~(2), and show them  
in the inset of Fig.~3(b). 
As expected for ordinary spin fluctuations, 
$H_{\Delta}$ exhibits almost no temperature dependence.
For $x$ = 1.0 and 0.95, $H_{\Delta} = 35$ G gives the simple
relation $\tau \sim 0.06 \lambda$ for 
$\tau$ in [$\mu$s] and $\lambda$ in [1/$\mu$s].  
Thus, the correlation time $\tau$ increases 
with decreasing temperature and keeps 
a constant value of $\tau$ = 10-20 ns 
below 1~K down to 50 mK. This suggests that dynamic  
Cu spin fluctuations persist below 1 K in a quantum 
(temperature independent) process.


Figure 4(a) shows the time spectra in pure and diluted samples  
in \LF = 100~G 
at 50~mK.    
The relaxation rate rapidly decreases with decreasing 
Cu concentration. For $x$=0.70 and 0.60, we observed  
almost no relaxation.   
We analyzed the time spectra with Eq.~(1). 
The $x$ dependence of $\lambda$ at 50 mK is shown in Fig.~4(b).
With increasing Zn doping,  
(decreasing $x$), $\lambda$ rapidly decreases down to $x$=0.90 
and then gradually decreases down to 0.60. 
Similarly to the case of $\chi$, the \mSR \ results vary smoothly
at the percolation threshold $p_{\rm c}$=0.65.
We could not estimate $H_{\Delta}$  
and $\tau$ for $x$=0.70 and 0.60, since $\lambda$ is almost 
independent of LF.  In the right hand axis of Fig. 4(b),
we show the correlation time $\tau$ at T = 50 mK, derived by
assuming $H_{\Delta}$ = 35 G. 
The Cu spin correlation time $\tau$ rapidly decreases 
with increasing dilution (decreasing $x$), indicating
faster spin fluctuations for less perfect spin network. 
%
\begin{figure}[h]
\label{fig:mSRe}
\input{epsf}
\epsfxsize 7.5cm
\centerline{\epsfbox{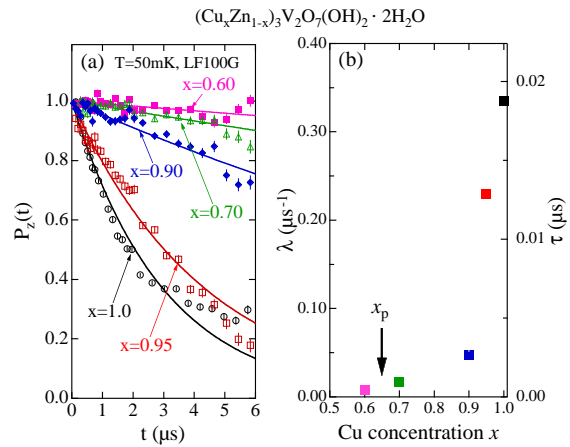}}
\caption{(a) Time spectrum in \LF = 100~G 
measured at 50~mK for various Cu concentrations $x$.  
(b) $x$ dependence of the relaxation rate 
$\lambda$ in \LF = 100~G measured at 50~mK 
and of the corresponding characteristic fluctuating time, $\tau$  
for $H_{\Delta}$ = 35 G.}
\end{figure}

Now let us consider the ground state of pure CVO. 
The absence of any oscillations in the $\mu$SR time spectra
down to 50~mK implies that there is no evidence for long-range order.
We found slow spin fluctuations, with the correlation time
of $\sim$ 10 ns, persisting below $T \sim 1$ K.  
The value of the instantaneous local 
field $H_{\Delta}$ for $x$=0.95 was nearly equal to 
that of $x$=1.0, while $H_{\Delta}$ decreases
with further Zn doping.   This suggests that 
the observed magnetism is not due to extrinsic states created 
by doped Zn$^{2+}$ ions via breaking of nearby singlet state(s) of Cu spins, 
but likely reflects an intrinsic 
property of the entire kagom\'e spin network.

In dilute spin glasses, the Lorentzian width of
internal fields due to dipolar (or other $1/r^{3}$)
interaction is inversely proportional to the average volume per
spin \cite{YJU85,Walstedt74}.  In CVO, the width $\sim 35$ G 
can be expected if 13 \% of the Cu spins have 0.5 $\mu_{B}$
moments pointing in random directions.  If these Cu spins
have antiferromagnetic correlations, one needs to assume
larger population of active Cu moments to explain 
the observed amplitude $H_{\Delta}$.  
Thus, the present results indicate that a significant
fraction of Cu moments remain active 
in the ground state, while
the rest of them may form magnetically inactive singlet states. 

In the present sample of nominally pure $x=1$ CVO, 
the $1/T$ impurity term of $\chi$  
correponds to $S$=1/2 free impurity of well below 0.25 \%\ 
of Cu population.  Thus the $\mu$SR results cannot be ascribed
to such impurity moments.  This point was further confirmed by
essentially identical $\mu$SR results obtained in another sample
of $x=1$ CVO which had nearly a factor of two larger $1/T$ term in 
$\chi$.

Some theories and numerical results 
suggested that the ground state of $S$=1/2 kagom\'{e} system is a 
singlet state associated with 
a rapid decrease of $\chi$ below $T \sim$0.2$J$~\cite{Sind00}. 
The broad maximum of $\chi$ observed 
in CVO at $T \sim 20$ K might be related to this feature. 
A similar broad maximum around $T \sim J/2$ was also found
in NMR spin susceptibility of SCGO~\cite{Mendels00}.
However, the ground states of both CVO and SCGO
involve active magnetism, as indicted by:
(1) the $\mu$SR relaxation in CVO (present study)
and SCGO~\cite{Uemura94}; (2) non-zero value of $\chi$ 
at $T\rightarrow 0$ even after the Curie-like term was subtracted;
and (3) quasi static response in SCGO at 
$T\rightarrow 0$ in neutron scattering studies~\cite{Broholm90, Uemura98}.
We note that we should not have seen {\em any} spin
relaxation at $T \rightarrow 0$ in a complete singlet ground state.
In both CVO and SCGO~\cite{Uemura94}, 
the $\mu$SR relaxation rates at low temperatures
are nearly independent of temperature.  These results
suggest a novel type of ground state in both $S$ = 1/2 (CVO) and
3/2 (SCGO) kagom\'e networks, where slow and quantum spin fluctuations
persist without static spin freezing.
The existence of such a slow magnetic state in CVO provides
a possible explanation for the broadening and disappearence
of an ESR signal in CVO below $ T \sim 4$ K~\cite{Okubo01}.
 
A spin-glass-like behavior
in $\chi$, associated with departure of field-cooled and zero-field 
cooled results, was found in the  kagom\'e systems 
SCGO~\cite{Ramirez90} and Cr-jarosite\cite{Keren96}. 
We point out that the \mSR\ results 
of CVO, SCGO~\cite{Uemura94,Keren00} and Cr-jarosite~\cite{Keren96}, 
with persisting
dynamic effects at $T\rightarrow 0$,  
are qualitatively different from those 
in ordinary spin-glass materials
such as $Cu$Mn~\cite{YJU85}, where dynamic fluctuations disapper at
$T\rightarrow 0$. 

Finally, we discuss the effect of magnetic dilution.
The present \mSR \ results in CVO show that the correlation time becomes 
rapidly shorter with increasing Zn doping.  Qualitatively similar
dependence on magnetic dilution was found in $\mu$SR
studies of SCGO~\cite{Keren00}.  These results indicate that
magnetic dilution does not relieve frustration
nor promote spin freezing in these systems.  Zn doping may cutoff the magnetic
network to smaller clusters resulting in faster spin fluctuations.
The absence of an anomaly at the percolation threshold,
in either CVO or SCGO, suggests that the 
spin fluctuations have a short-ranged / local character.  

In summary, we found a signature of slowing down of 
spin fluctuations with decreasing 
temperature, followed by 
slow and nearly temperature-independent
spin fluctuations between $T$ = 1.0 and 50 mK 
in the $S$=1/2 kagom\'e spin system CVO.
These fluctuations exhibit  
quantum character, and persist
at $T$ = 50 mK without any signature of long-range or static
spin order.  

The work at Columbia has been supported by the National Science Foundation 
DMR-0102752 and CHE-0117752 (Nanoscale Science and Engineering
Initiative).  Research at McMaster is supported by
NSERC and the CIAR (Quantum Materials Program).

%







\end{document}